\documentclass[letter]{aa} 

\usepackage[dvips]{graphicx}
\usepackage{txfonts}
\usepackage[usenames]{color} 
\usepackage{natbib}

\newcommand{\modif}[1]{#1}
\newcommand{\spellcheck}[1]{#1}
\newcommand{\edspell}[1]{}

\begin{document}

\title{Polar plumes' orientation and the Sun's \spellcheck{global} magnetic field}

\author{Judith de Patoul\inst{1,2} 
     \and
	Bernd Inhester\inst{1}
     \and
	Robert Cameron\inst{1}  
}
\institute{Max Planck Institute for Solar System Research,
	Max-Planck-Str. 2, 37191 Katlenburg-Lindau, Germany\\
	\email{depatoul@mps.mpg.de}, 
	\email{inhester@mps.mpg.de},
	\email{cameron@mps.mpg.de}
     \and
        Laboratoire de Physique des Plasmas,
        \'Ecole Polytechnique, CNRS, 91128 Palaiseau Cedex, France\\
	\email{judith.de-patoul@lpp.polytechnique.fr}
}

\date{Today:\today ; Received ; accepted}


\abstract
{}
{We characterize the orientation of polar plumes as a tracer of the
large-scale coronal magnetic field configuration. We monitor in particular the
north and south magnetic pole locations and the magnetic opening during
2007--2008 and provide some understanding of the variations in these
quantities.}
{The polar plume orientation is determined by applying the Hough-wavelet
transform to a series of EUV images and extracting the key Hough space
parameters of the resulting maps. The same procedure is applied to the polar
cap field inclination derived from extrapolating magnetograms generated
by a surface flux transport model.}
{We observe that the position where the magnetic field is radial (the Sun's
magnetic poles) reflects the \spellcheck{global} organization of magnetic field on the
solar surface, and we suggest that this opens the possibility of both
detecting flux emergence anywhere on the solar surface (including the far
side) and better constraining the reorganization of the
corona after flux emergence.}


\maketitle


\section{Introduction}
\label{sec_intro}

The plasma $\beta$ in the corona is low enough that, on time scales much
greater than the wave-crossing times and on spatial scales much greater than
the small-scale turbulence driven by photospheric braiding of the field lines,
it is almost force free. The large-scale, slowly-evolving coronal magnetic
field is then predominantly determined by the distribution of the magnetic field
in the photosphere below and by the way it connects to the solar wind
and parker spiral above. This is the basis for attempts to reconstruct the
coronal magnetic field using photospheric vector magnetograms \citep[for a
review see][]{Wiegelmann12}. Additional information on the spatial structure
of the coronal magnetic field is contained in EUV and X-ray images of the
solar corona \cite[see][]{Aschwanden08}, which can be used to test
field extrapolations \citep[see][]{DeRosa09}.

For \spellcheck{global} extrapolations of the Sun's magnetic field \cite[for a recent
review see][]{Mackay12}, the total amount of open flux \cite{Lockwood04}, the location
of the heliospheric current sheet \citep[e.g.,][]{Virtanen2010, Erdos2010}, and
the location of coronal holes \cite[e.g.,][]{Levine82} are additional
constraints.
For example, \cite{Wang97} studied the
evolution of the coronal streamer belt during the 1996 activity
minimum and compared the
results to extrapolations, including those based on the evolution of the
photospheric magnetic field from a flux transport simulation \cite[see,
e.g.,][]{Mackay12}. They found that changes in the structure of the
streamers correspond to flux emergence, in such a way that both the longitude
of emergence and the orientation of the newly emerged bipole to the
pre-existing coronal fields affect the streamer's structure.

In this paper we discuss a further constraint for models of the coronal
magnetic field, which is induced by the orientation of polar plumes,
predominantly elongated striations seen in coronagraphs mostly above the polar
caps. The orientation of these plumes is aligned with the magnetic field and
can be directly compared with the orientation of field lines from
extrapolations of the coronal magnetic field.
We perform an analysis related to the work of \cite{Wang97} using EUV
images from SOHO and concentrating on polar plumes. This restricts our study
to times of solar activity minimum, because these are the times when the
plumes are readily visible.
Here, we use five parameters to characterize
the orientation of all plumes visible in one hemisphere at any given time. The
first two parameters are the longitude and latitude at which the inclination
of the plumes is exactly radial: on the assumption that the plumes are field-aligned
this corresponds to the magnetic north (or south) pole.
The three other parameters characterize how rapidly the
plumes tilt away from radial with increasing distance from the pole
as a function of longitude.
These parameters measure the magnetic \modif{opening factor} of the polar cap
field.
To lowest order, we can determine two such  \modif{opening factors} in
mutually orthogonal magnetic meridional planes and the longitudinal
orientation of these planes. Specific details, including how these parameters
were measured, are given in Section 2.

In Section 3 we give the time history of the evolution of the location of the
magnetic north and south poles, as well as the  \modif{opening factors}. This evolution
is compared with current sheet source surface extrapolations \citep{Zhao95}
based on surface flux transport simulations\citep{Jiang2010}.
We find reasonable qualitative agreement, with the differences suggesting
ways the surface flux transport model can be refined.
In Section 4 we note that the orientation of the coronal plumes is thus
sensitive to flux emergence anywhere on the Sun \citep[as][found for
streamers]{Wang97}. As with the streamers, the temporal variation of the
plume orientation is sensitive to flux emergence anywhere on the disk.
The changes depend on where the flux
emerges, how it is oriented in relation to the pre-existing field, and how
the field of the bipole and the pre-existing field interact. Although not
pursued in this paper, this suggests that ``far-side imaging'' of flux emergence
and studies of the interaction between the emerging flux and pre-existing
field using coronal images is a possibility. Section 5 briefly presents our
conclusions.


\section{Methods}
\subsection{Determination of the polar coronal field inclination }
\label{sec_meth}
The raw data for this study are sequences of EUVI and EIT images at 171~{\AA}
taken from April 3 2007 to May 8, 2008, i.e., during the minimum between
cycles 23 and 24. Our aim is to use the orientation of the polar plumes as a
diagnostic of the Sun's large-scale magnetic field. The procedure is to first
analyze individual images. For the STEREO/EUVI and SOHO/EIT 171\AA\ images, we have
calculate the Hough-wavelet transform as described in \cite{dePatoul13} 
to measure the inclination of the plumes.

Briefly, the images have been preprocessed by suitable background subtraction and
mapped into cylindrical coordinates $(r,\phi)$ with respect to the solar
disk center. The images have then been transformed using a Hough-wavelet transform to a
translational distance ($\rho$) and an inclination ($\theta$).

As is shown in \cite{dePatoul13}, power maxima in the transformed map are
generated at singular positions ($\rho,\theta$) when the projected plume
structure is inclined by angle $\theta$ with respect to the radial, and its
foot point is projected to colatitude $\vartheta_c = \rho/\cos\theta-\Delta
\tan\theta$. Here, $\Delta$ is the distance of a virtual Hough center above the
solar surface in units of solar radii.
Since the inclination angle $\theta$ of the plumes varies smoothly with colatitude
$\vartheta_c$, or likewise with $\rho$, the power maxima lie in Hough space
almost along a straight line $\theta = u \rho + v$.
The $\rho$-value where this line crosses $\theta=0$ (intercept $-v/u$) gives
the special colatitude $\tilde\vartheta_c$ of the foot-point from which
a plume is projected perpendicular to the limb.
The slope $u$ of this line corresponds to the change in the plume's
inclination $\iota$ with foot-point colatitude $\vartheta_c$. Introducing the
co-inclination angle $\iota_c=\pi/2-\iota$ which vanishes at the pole, the
parameter $u$ is, to lowest order, directly proportional to the opening factor
$f_c=\tan\iota_c/\tan\vartheta_c$ at $\theta_c=\tilde\theta_c$. For a perfect
dipole field, this factor has the value of 0.5.%

The result for each image is thus two parameters, the intercept $-v/u$ and the
slope $u$. For each spacecraft, we have a sequence of images so that
we obtain a time series for these two parameters. They are shown in
Figs.~\ref{fig_ts_data}, where the time series from the STEREO spacecraft
(which have different view directions onto the Sun) have been corrected
by a time shift according to the spacecraft's angular
distance from Earth so as to make the time series comparable to the
SOHO/EIT viewpoint. The results from the different space
craft data are indicated by different colors. Where they overlap \spellcheck{well}, the
residual scatter in the parameters is due to image noise. Enhanced scatter
reveals intrinsic changes in the polar cap plume configuration.

In principle we could have used these projection parameters
from simultaneous observations of the SOHO and the two STEREO spacecrafts to
reconstruct the instantaneous three-dimensional pole position on the surface
by stereoscopy.
This would have limited the method, however, to times when observations from
multiple vantage points are available. To make this method wider
applicability, we instead maked the assumption that, for most of the time, the
magnetic field does not change rapidly compared with the solar rotation. This
allows us to use the solar rotation to obtain different projections on the
plane of the sky and to derive the three-dimensional parameters from the time
variation of their projections. Infrequent rapid time changes in the
projections that hint at emergence events will show up as sudden changes in
the measured parameters, and so are readily visible.

\subsection{Determination of the solar poles}

\begin{figure*}
\centering
\includegraphics[width=0.9\textwidth, trim=0.6cm 0 0.6cm 0,clip]{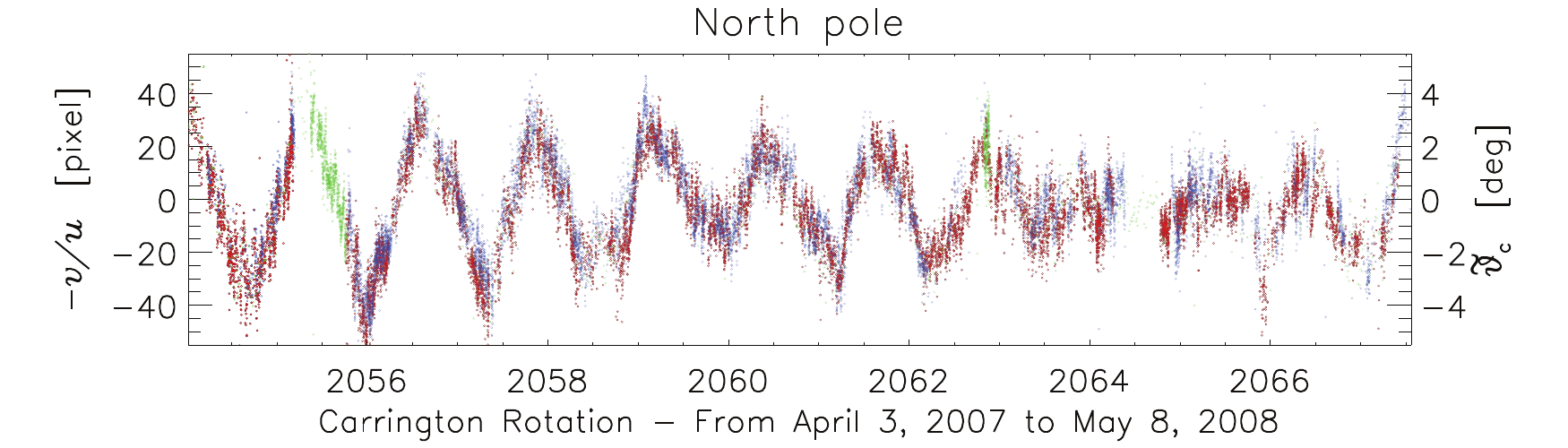} \\
(a)\\
\includegraphics[width=0.9\textwidth, trim=0.6cm 0 0.6cm 0,clip]{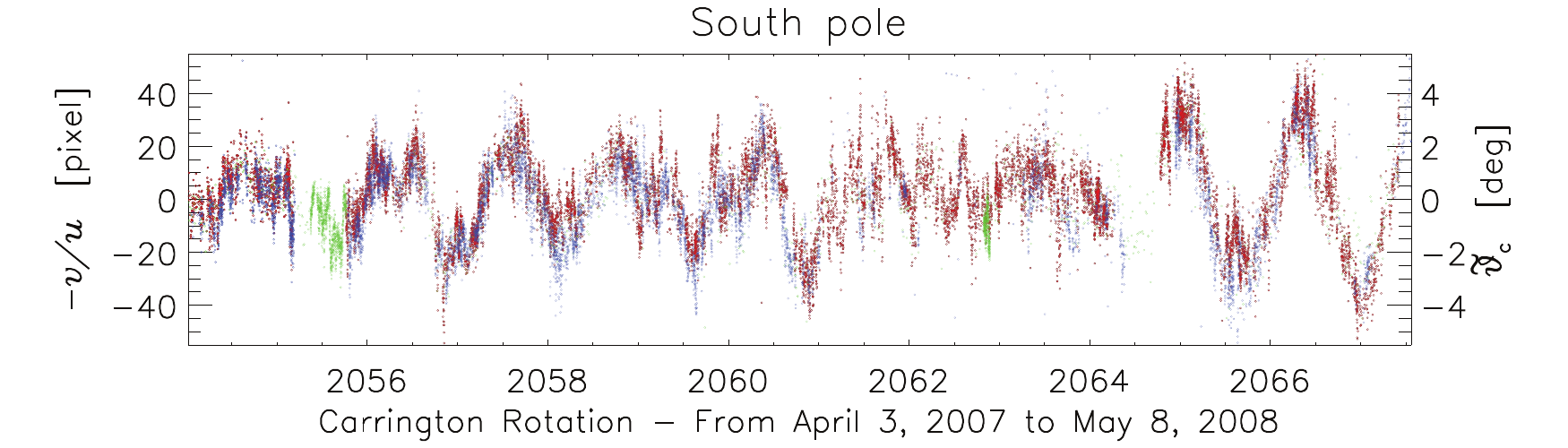} \\
(b) \\
\includegraphics[width=0.9\textwidth, trim=0.1cm 0 0.6cm 0,clip]{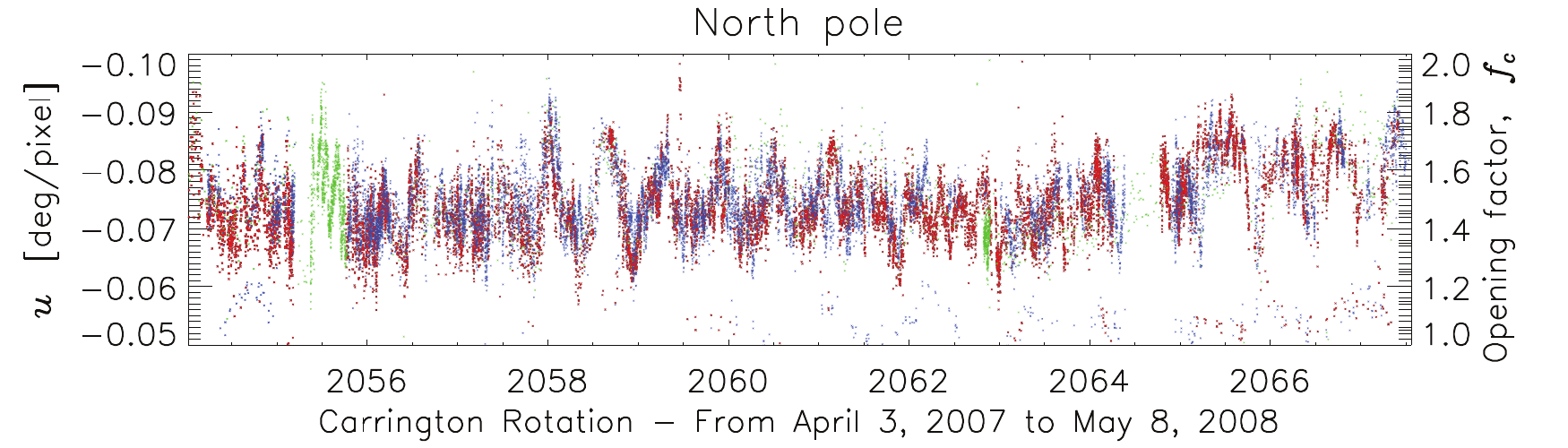} \\
(c) \\
\includegraphics[width=0.9\textwidth, trim=0.1cm 0 0.6cm 0,clip]{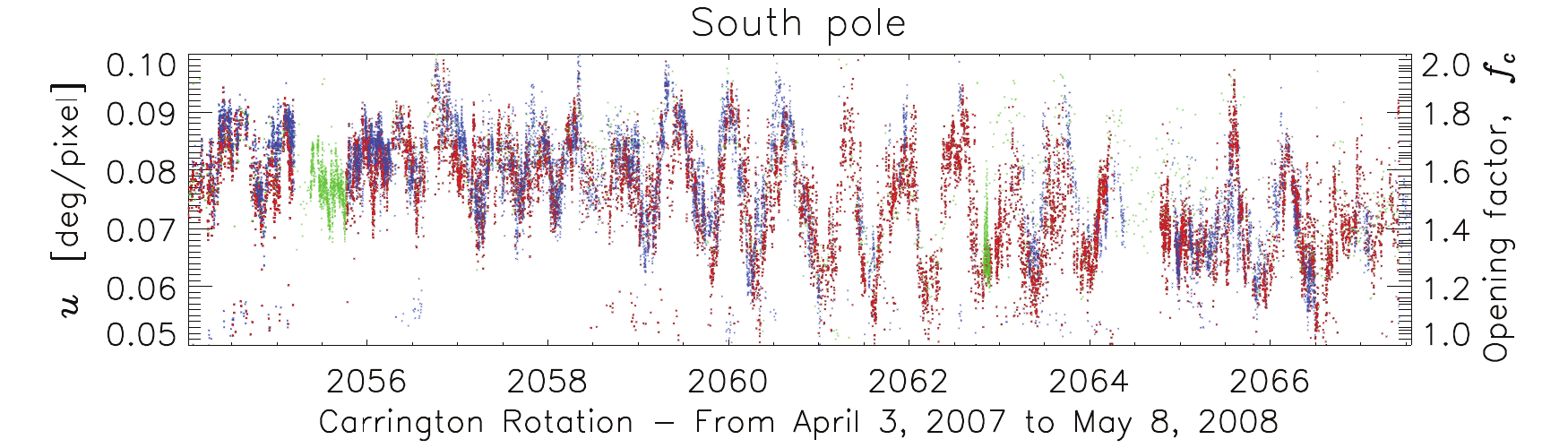} \\
(d) \\
\caption{Time series of the intercept $-v/u$ (a and b) and the slope parameter
$u$ (c and d) calculated for the north and south poles from data of EUVI~A (in
red), EUVI~B (in blue), and EIT (in green) in the wavelength of 171~\AA.
The contributions of the different instruments are indicated by different
colors.
As explained in the text, the intercept is proportional to the projected
heliographic colatitude $\tilde\vartheta_c$ of the magnetic pole location. The
slope is a measure of the projected opening factor $f_c$ of the polar magnetic
field at the respective pole (see labeling of the respective right ordinate
axes).
}
\label{fig_ts_data}%
\end{figure*}

When the surface field only varies slowly, the time dependence of the
intercept $-v/u$ will be dominated by the Sun's rotation.
The intercept value
should then oscillate with the Sun's rotation frequency and with an amplitude that
corresponds to the colatitude of the pole's position. Its longitude is
related to the phase of this oscillation: at the time when the
intercept parameter is zero and rising, the pole is tilted towards Earth.
To lowest order, this is what we observe in the recordings of the
intercept $-v/u$ in Figs.~\ref{fig_ts_data}a and b.
Smooth changes in amplitude and phase indicate gentle motions
of the pole (secular variation), and rapid changes allude to a rapid
reorganization of the polar and probably of the whole coronal field. For these
latter instances, our determination of the pole position is not rigorously
valid. For the time period studied, the intercept parameter showed a mean period
of $34.09$ days, and the location of the poles derived from
an instantaneous oscillation amplitude and phase is shown in the upper panel
of Fig.~\ref{fig_pole_data_Tobs}.

Similarly, the time variation of the slope parameter displays
an oscillation with nearly twice the solar rotation frequency with a mean
offset (see Fig.~\ref{fig_ts_data} c and d). Again for a stationary field, the minima and
maxima of this oscillation are related to the minimum and maximum \modif{opening factors}
of the coronal pole field. To lowest approximation, these \modif{opening factors} are defined
in two orthogonal magnetically meridional planes that, except for a tilt of
the magnetic pole towards or away from the observer, come close to the
observer's plane of the sky at times when the slope parameter assumes a
maximum or a minimum.

\subsection{Surface flux transport modeling and current sheet source surface
  model}

\begin{figure*}
\centering
\includegraphics[width=0.9\textwidth]{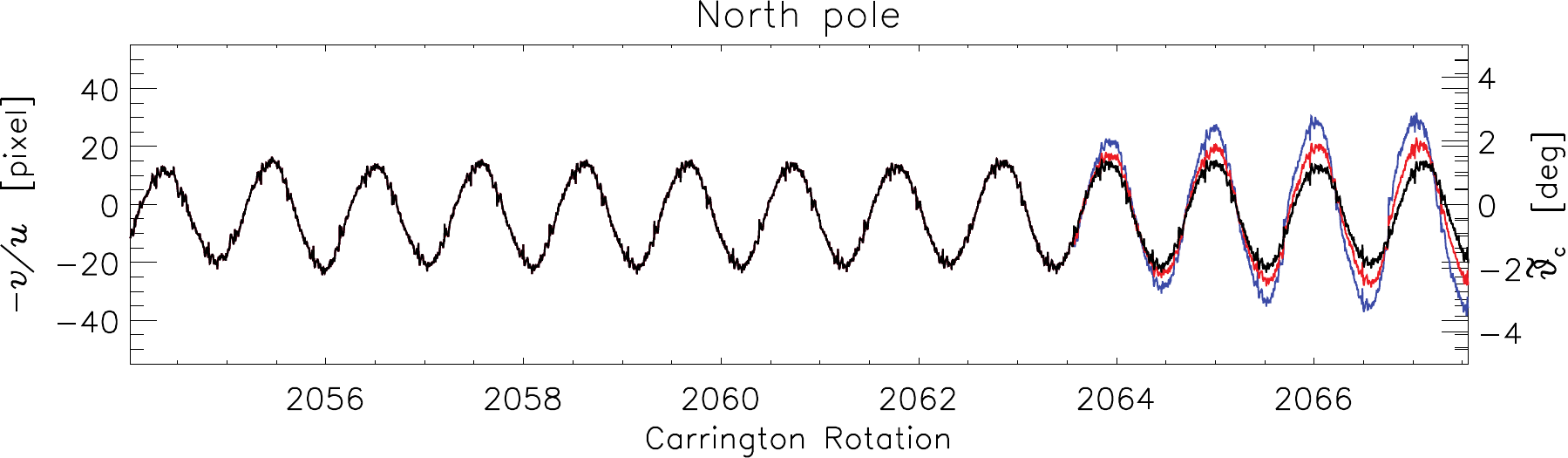}\\
\includegraphics[width=0.9\textwidth]{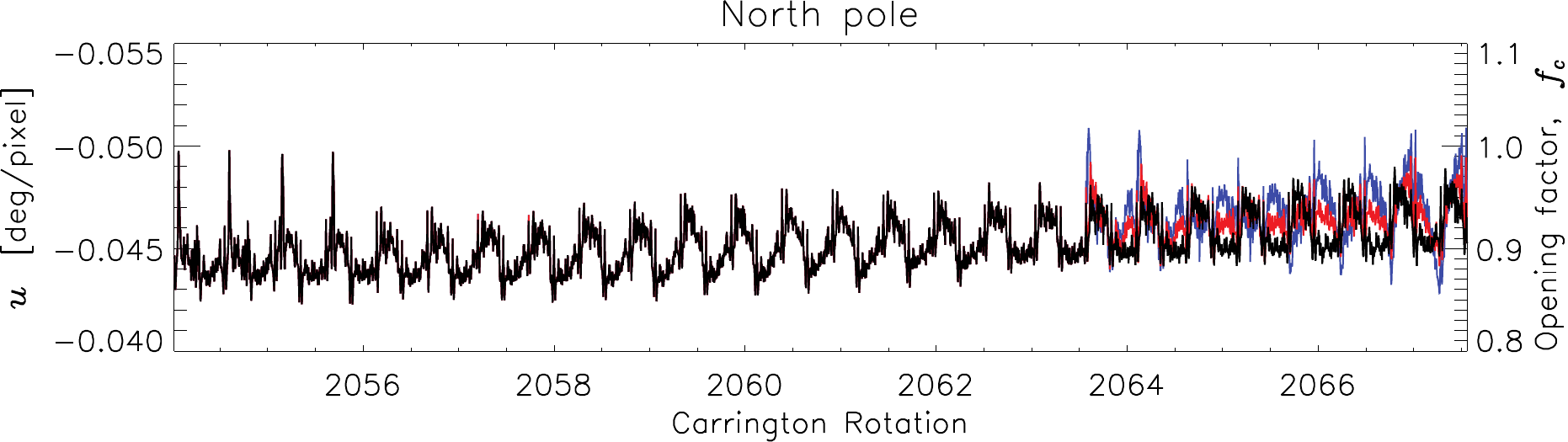}
\caption{Time series of the intercept $-v/u$ (upper panel) and the \modif{opening factor} 
parameter $u$ (bottom panel) as in fig.~\ref{fig_ts_data} but calculated from the
surface flux transport/current sheet source surface model \modif{(in black). The blue and 
red curves are the results from the model with extra sunspot groups 
corresponding to the synoptic magnetograms in Fig.~4, upper and middle panels, respectively.}
We only display the results for the northern hemisphere here. }
\label{fig_ts_simul}%
\end{figure*}

\begin{figure*}
\centering
\includegraphics[width=0.75\textwidth]{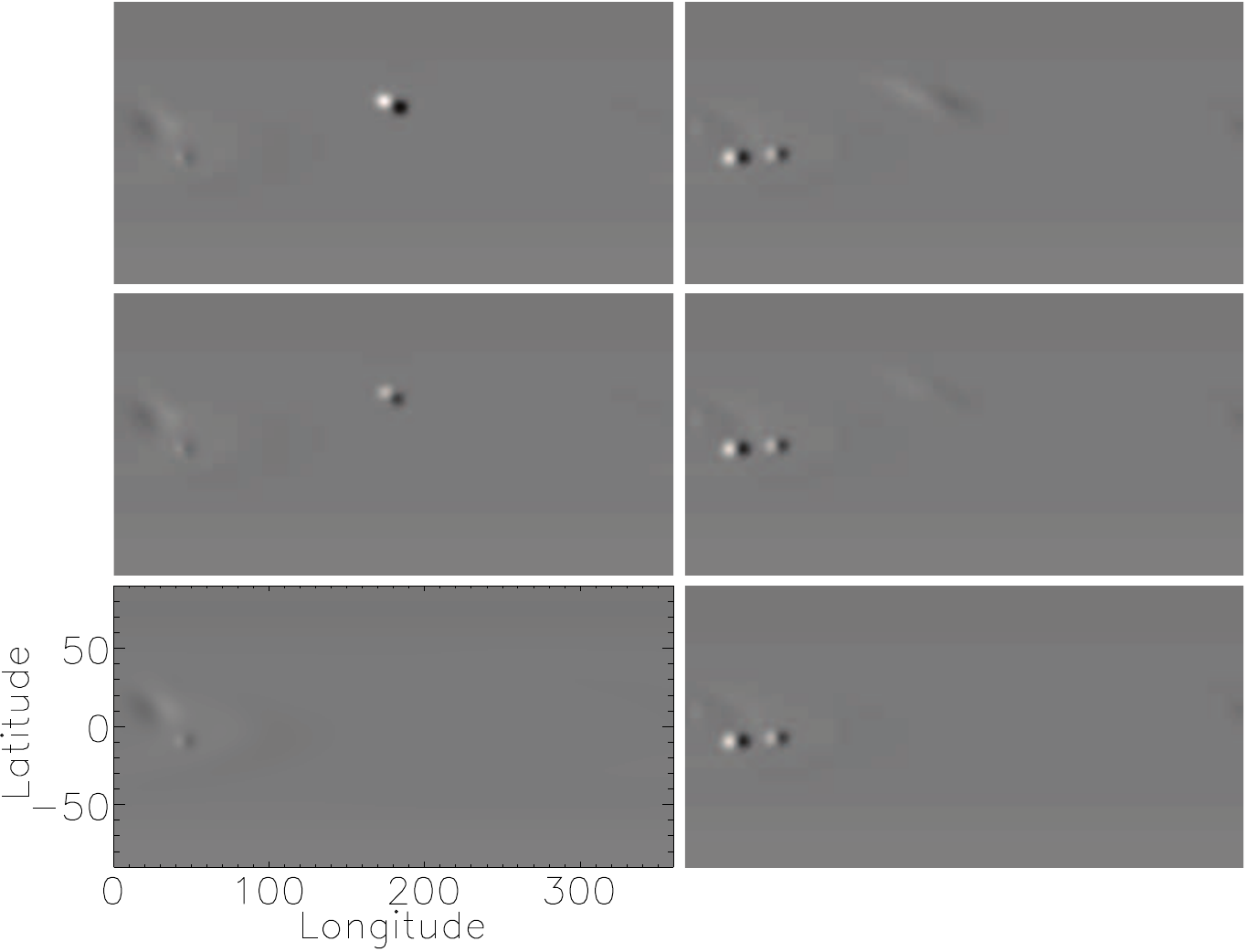}
\caption{\modif{
Original synoptic magnetograms (bottom row) and two simulated
magnetograms (top and central rows) with artificial sunspots
of different strength. The left panels show the magnetograms
with the sunspots at the time of their emergence, the right
panels the magnetograms 54 days after emergence.
}}
\label{fig_manetograms}
\end{figure*}

As mentioned in the introduction, as an illustrative application of the
method, we compare the observations with the magnetic field derived from
surface flux transport simulations and a current sheet source surface
extrapolation at each time step. For simplicity we used the results described in detail in
\citep{Jiang2010} and references therein. The source term for these
simulations is based on the recorded sunspot records \citep[again see][for
details]{Jiang2010}. Since the observations were taken during activity minima,
relatively few sunspots were recorded for this period. From this
model we extracted the field orientation at $1.19$~R$_{\odot}$ and on the
plane of the sky. Similar to the processing of the plume orientations,
the projected field orientations were transformed into cylindrical coordinates 
$(r,\phi)$ with respect to the solar disk center and subsequently processed
by the  Hough-wavelet transform. The resulting time series for the same
time period covered by the observations is shown in Fig.~\ref{fig_ts_simul}.
The solar rotation period derived from the simulated data was 28.67 days.
\modif{To test how the magnetic pole location is sensitive to the emergence of sunspots, 
we artificially introduce a sunspot group at 2064 Carrington rotation. Two cases of sunspots  are considered
with different strengths and with same location. Their respective magnetograms are shown in 
Fig.~\ref{fig_manetograms}, where we can see that the artificial sunspot groups 
are comparable to the other groups that were present 
at that time. The magnetic pole location and the opening factor are calculated 
and shown in Fig.~\ref{fig_ts_simul}. 
The location of the poles is derived and shown in Fig.~\ref{fig_pole_data_Tobs} (bottom panel).}

\section{Discussion}
\label{sec_resu}

\begin{figure*}
\centering
\includegraphics[width=0.6\textwidth]{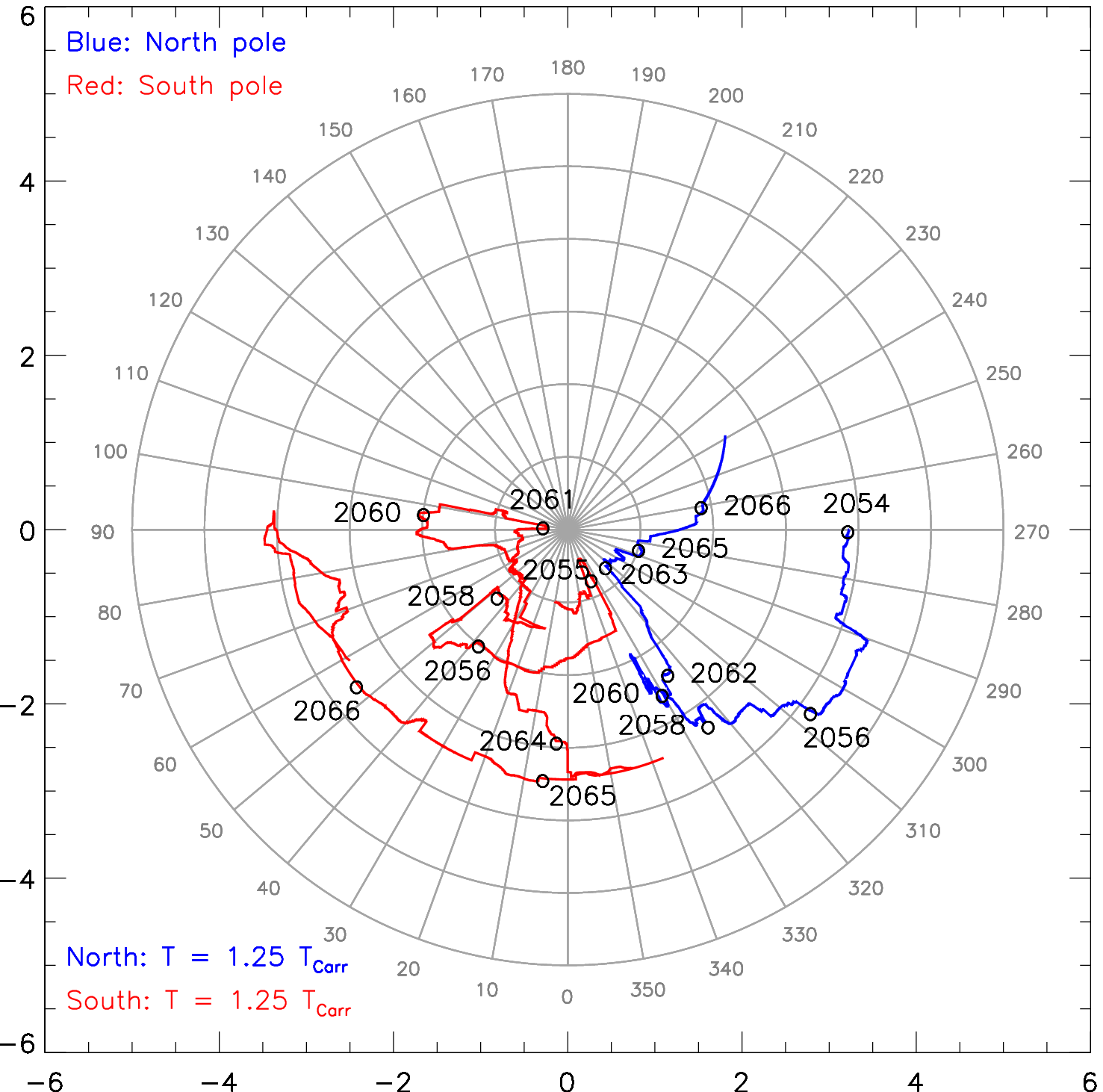} \\
\includegraphics[width=0.6\textwidth]{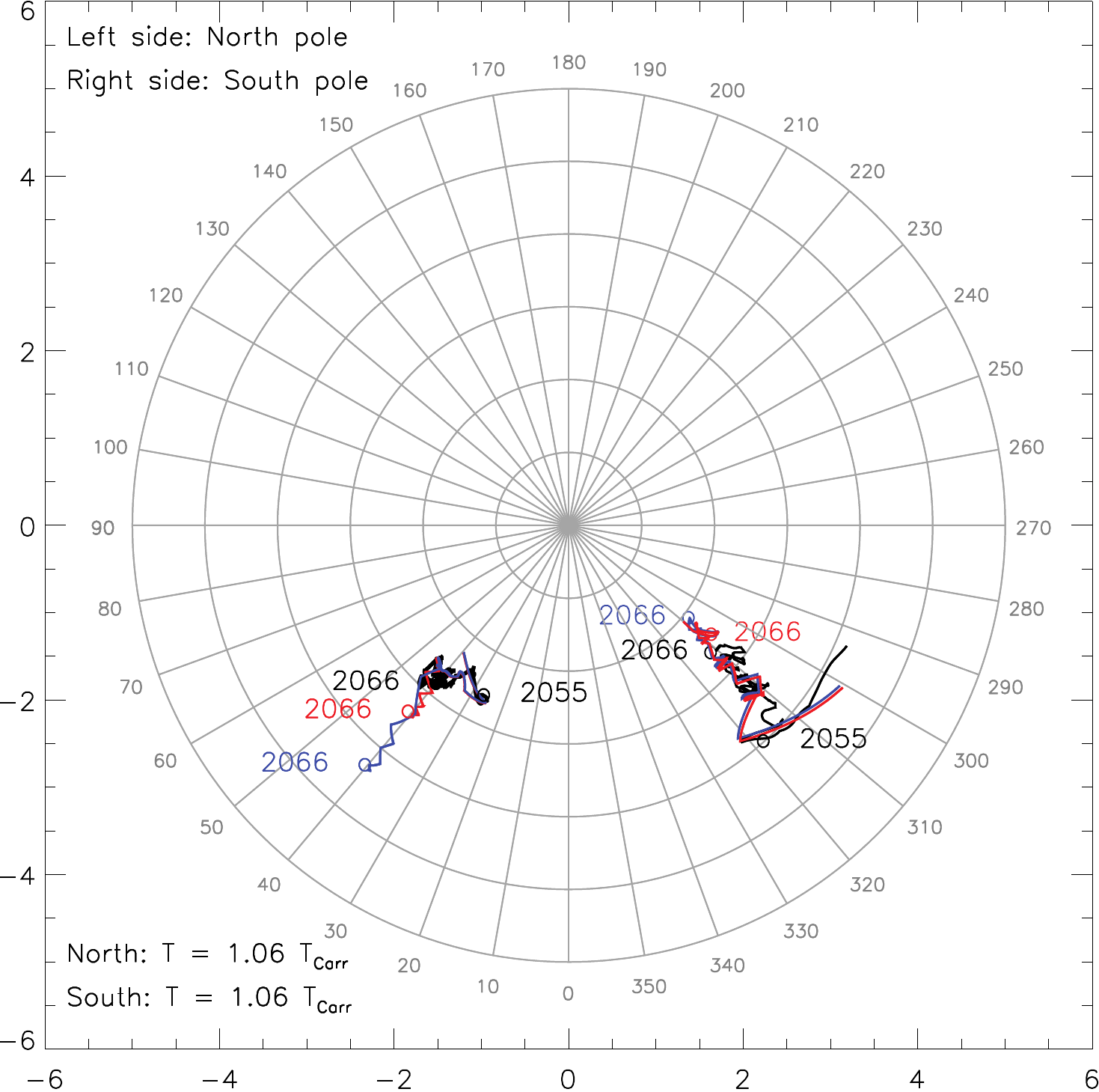} 
\caption{The evolution of the location of the Sun's magnetic poles in a
reference frame corotating with respect to the polar cap.
North and south poles are drawn in blue and red, respectively.
The concentric circles mark colatitudes from the heliographic pole
in 1 degree steps.
In the upper panel, the pole path is derived from the parameters displayed
in Fig.~\ref{fig_ts_data}.
The rotational period that eliminates the average longitudinal drift of the 
north pole is 34.09 days.
The lower panel shows the same evolution based on a surface flux transport
simulation and a current sheet source surface extrapolation, \modif{in black}. 
The rotational period of the nonaxisymmetric field was found in the simulation
to be 28.67 days. \modif{The red and blue lines correspond to the model that includes
artificial sunspots as discussed in the text.}}
\label{fig_pole_data_Tobs}
\end{figure*}

The evolution of the magnetic poles position in the simulated case (bottom panel of 
Fig.~\ref{fig_pole_data_Tobs}, \modif{in black}) is particularly simple: 
the poles rotate at a rate of 28.67 days and 
slowly drift towards the rotational pole (especially apparent in the southern hemisphere). 
This simple evolution reflects that no 
sunspots emerged during this time and so that no newly 
emerged flux were included for this time period in the simulation. 
The slow drift of the magnetic pole towards
the rotational pole is due to the decay of a previously existing non-axisymmetric field 
as it is dispersed over all longitudes by differential rotation.
\modif{When artificial sunspot groups are included in the simulation, the results
show a clear change in the $-u/v$ parameter in Fig.~\ref{fig_ts_simul} and in 
the magnetic pole locations in Fig.~\ref{fig_pole_data_Tobs}, which 
demonstrates that the magnetic pole location is sensitive to the flux emergence 
anywhere on the solar surface.}

The evolution of the magnetic pole derived from the plume observations is more
complicated. First, the the rotational period of 28.67 days is only a mean
value, and there are times when the pole rotates faster and times when it
rotates slower. This, in part, reflects that the rotation rate of
magnetic fields do vary in time \citep{Singh85}. Second, there is a
more erratic behavior than in the simulations, with abrupt changes in the
latitude of the poles. This directly reflects the amplitude
changes of the oscillation of the
intercept $-v/u$ seen in the left hand panels of Fig.~\ref{fig_ts_data}. As in \cite{Wang97},
these changes in the coronal magnetic field are the result of small-scale flux
emergences, which modifies the \spellcheck{global} magnetic field of the entire corona  
within an Alfv\'en transit time scale. An
examination of SOLIS magnetograms covering this period showed, as 
expected, several small-scale flux emergence events that did not produce
sunspots and therefore were not included in the surface flux transport
simulations. This explains why the model intercept parameter $-v/u$ has a much cleaner
periodic signal than the one based on the observations. On the other hand, it demonstrates the 
sensitivity of the parameters derived here to the structure of the coronal field.

Newly emerging flux, wherever it occurs on the Sun, produces a signature
in the coronal magnetic field \cite{Wang97} and, in particular, in the
orientation of the polar plumes.
Presently, the imaging of flux emergence on the far side of the Sun is
made possible by complex inversion calculations of front-side
helioseismology data \citep{Lindsey00}.
Although not pursued in this paper, our study suggests that a continuous monitoring of
the orientation of polar plumes might improve our capability of 
far-side imaging of flux emergence.

The other parameters we measured were the \modif{opening factors}.
\spellcheck{Their} average corresponds to the mean slope parameter $<u>$, which
reflects how much the polar field spreads out with distance from the solar
surface.
This parameter is about 50\% larger in our observations than in the simulations. This
indicates that the observed polar fields expand more rapidly from the
photosphere than our simulations predict. 
\modif{A possible explanation for the discrepancy} \spellcheck{is a higher} concentration of the photospheric field towards the
poles of the Sun.
The enhanced concentration suggests that the
meridional velocity on the Sun is more efficient in
convecting the magnetic field to the (rotational) poles
than assumed in our flux transport simulations.
These velocities are indeed poorly constrained by
observations and in the flux transport simulations used here \citep{Jiang2010}.
The meridional flow was simply assumed to vanish at \modif{15 degrees} from
the poles. In this regard the observed average \modif{opening factor}
provides a sensitive constraint on the meridional flow near
the poles where it is difficult to measure.

\section{Conclusion and summary}
\label{sec_ccl}
We have carefully characterized the orientation of polar plumes and
derived the time-dependent locations of the north and south
magnetic poles.
By comparing these results against surface flux transport simulations
coupled with current sheet source surface models, we found that these
properties are sensitive to the \spellcheck{global} magnetic field configuration, so
are potentially useful in both understanding how the large-scale field
reorganizes after flux emergence, and constraining the properties of
flux emergence on the far side of the Sun. The average  \modif{opening factor}
of the polar fields is suggested as a constraint on the strength of the
near-polar meridional flow, which should be considered in future studies.

Our study supplements the earlier work of \cite{Wang97}, who
used the orientation of streamers above the coronal belt to monitor
reconfigurations of the coronal magnetic field. Using the orientation of
plumes instead has the disadvantage that \modif{they are only observed during 
the solar activity minimum.} 
Since flux emergence events occur most often
at low and middle latitudes, the polar cap field is modified less by
these distant sources, and the polar cap field 
sees \spellcheck{more of an integral} effect rather than local
emergences. 
Definite advantages of our approach is that it is readily
automated, it is continuously sensitive to all longitudes, and it can distinguish
between temporal and spatial changes.

\begin{acknowledgements}
SOHO and STEREO are joint projects of ESA and NASA.
The analysis of the STEREO data was performed with support from DLR
grants 50OC0501 and 50OC1301
\end{acknowledgements}

\bibliographystyle{aa}
\bibliography{article.bib}  

\end{document}